%% file: main.tex
\documentclass[10pt,twocolumn,letterpaper]{article}

\usepackage{cvpr}              

\input{preamble}

\definecolor{cvprblue}{rgb}{0.21,0.49,0.74}
\usepackage[pagebackref,breaklinks,colorlinks,allcolors=cvprblue]{hyperref}

\def\paperID{*****} 
\def\confName{CVPR}
\def\confYear{2026}

\title{Halfway to 3D: Ensembling 2.5D and 3D Models for Robust COVID-19 CT Diagnosis}

\author{
Tuan-Anh Yang$^{1}$ \quad
Bao V. Q. Bui$^{2}$ \quad
Chanh-Quang Vo-Van$^{1}$ \quad
Truong-Son Hy$^{3}$ \\
\\
$^{1}$VNUHCM University of Science, Vietnam National University, Vietnam \\
$^{2}$Ho Chi Minh University of Technology, Vietnam National University, Vietnam \\
$^{3}$The University of Alabama at Birmingham, United States \\
\\
{\tt\small ytanh21@apcs.fitus.edu.vn, vvcquang19@clc.fitus.edu.vn, bao.bui171@hcmut.edu.vn, thy@uab.edu}
}

\begin{document}
\maketitle
\input{sec/0_abstract}    
\input{sec/1_intro}
\input{sec/2_related_work}
\input{sec/3_methodology}
\input{sec/4_experiments}
\input{sec/5_conclusion}
{
    \small
    \bibliographystyle{ieeenat_fullname}
    \bibliography{main}
}


\end{document}

%% file: preamble.tex
\usepackage{makecell}

%% file: sec/0_abstract.tex
\begin{abstract}
We propose a deep learning framework for COVID-19 detection and disease classification from chest CT scans that integrates both 2.5D and 3D representations to capture complementary slice-level and volumetric information. The 2.5D branch processes multi-view CT slices (axial, coronal, sagittal) using a DINOv3 vision transformer, while the 3D branch employs a ResNet-18 architecture pretrained with Variance Risk Extrapolation (VREx) and further refined with supervised contrastive learning to improve cross-source robustness. Predictions from both branches are combined via logit-level ensembling. 

Experiments on the PHAROS-AIF-MIH benchmark demonstrate the effectiveness of the proposed approach. On the test set, our method achieved \textbf{runner-up} in the Multi-Source COVID-19 Detection Challenge, with the best ensemble reaching a Macro F1-score of \textbf{0.751}. For the Fair Disease Diagnosis Challenge, our approach ranked \textbf{third place}, achieving a best Macro F1-score of \textbf{0.633} with improved performance balance across genders. These results highlight the benefits of combining pretrained slice-based representations with volumetric modeling, as well as the importance of ensemble strategies for improving robustness and fairness in multi-source medical imaging tasks. Code is available at \url{https://github.com/HySonLab/PHAROS-AIF-MIH}.
\end{abstract}

%% file: sec/1_intro.tex
\section{Introduction}
\label{sec:intro}

Chest computed tomography (CT) imaging plays a key role in the diagnosis and assessment of COVID-19 and other pulmonary diseases. Compared to conventional radiography, CT scans provide detailed volumetric information that enables the identification of abnormalities such as ground-glass opacities and lung consolidations. With the growing availability of medical imaging data, deep learning methods have shown strong potential for automated disease detection and analysis from CT scans \cite{kollias2021mia,kollias2023ai}.

Despite these advances, developing robust medical imaging models remains challenging. Clinical datasets are often collected from multiple institutions with different scanners, acquisition protocols, and patient populations, leading to domain shifts that degrade model generalization. In addition, fairness has become an important concern in medical AI, as models may exhibit performance disparities across demographics. Benchmarks such as the PHAROS-AIF-MIH challenge emphasize the need for methods that are robust to heterogeneous data sources while maintaining reliable performance across patient populations \cite{kollias2025pharos}.


Our method consists of two complementary branches. The 3D branch employs a ResNet-18 architecture trained with Variance Risk Extrapolation (VREx) to improve domain generalization, followed by supervised contrastive learning for enhanced feature discrimination. In parallel, the 2.5D branch extracts axial, coronal, and sagittal slices from reconstructed CT volumes and processes them using a DINOv3-based vision transformer. The predictions from both branches are combined through an ensemble strategy to leverage both slice-level and volumetric information.

We evaluate the proposed framework on the PHAROS-AIF-MIH benchmark for multi-source disease diagnosis and fairness-aware evaluation. Experimental results demonstrate that integrating 2.5D and 3D representations improves robustness across heterogeneous imaging sources while maintaining strong diagnostic performance.

The main contributions of this work are as follows:

\begin{itemize} 
\item We propose a hybrid 2.5D–3D framework that combines multi-view slice representations with volumetric modeling for CT-based diagnosis. 
\item We incorporate domain generalization using VREx and supervised contrastive learning in the 3D branch. 
\item We design a multi-view 2.5D pipeline using a DINOv3 backbone to leverage pretrained visual representations. 
\item We show that ensembling 2.5D and 3D models improves robustness and performance on the PHAROS-AIF-MIH benchmark. 
\end{itemize}

%% file: sec/2_related_work.tex
\section{Related Work}
\label{sec:related_work}

Deep learning has been widely applied to medical imaging for automated disease diagnosis. Early work demonstrated that neural networks can effectively extract diagnostic features from medical images, enabling computer-aided decision support in healthcare \cite{kollias2018deep}. Subsequent studies emphasized the importance of interpretability and reliability in medical AI systems, introducing techniques for analyzing latent representations and improving the transparency of deep learning predictions in clinical settings \cite{kollias2020deep,kollias2020transparent}.

For COVID-19 diagnosis, several studies have explored chest CT scans for automated detection and severity assessment. The MIA-COV19D framework introduced a large-scale dataset and deep learning pipeline for COVID-19 detection from 3D CT volumes \cite{kollias2021mia}. Later works proposed improved architectures, datasets, and AI pipelines to enhance model robustness and scalability for CT-based diagnosis \cite{arsenos2022large,kollias2022ai,kollias2023deep,kollias2023ai,gerogiannis2024covid}. More recently, multimodal and vision-language approaches such as SAM2CLIP2SAM have been introduced to further improve CT scan understanding \cite{kollias2024sam2clip2sam}.

To address the aforementioned domain shifts inherent in multi-institution datasets like the PHAROS benchmark, domain generalization methods such as Variance Risk Extrapolation~\cite{yuan2025multi} enforce domain-invariant representations.

The PHAROS-AIF-MIH benchmark further emphasizes the importance of multi-source robustness and fairness-aware evaluation in medical AI systems \cite{kollias2025pharos}. Motivated by these challenges, we propose a hybrid framework that combines 2.5D multi-view representations and 3D volumetric modeling to improve robustness and diagnostic performance for CT-based disease classification.

%% file: sec/3_methodology.tex
\section{Methodology}
\label{sec:methodology}


\begin{figure*}[h]
    \centering
    \includegraphics[width=\linewidth]{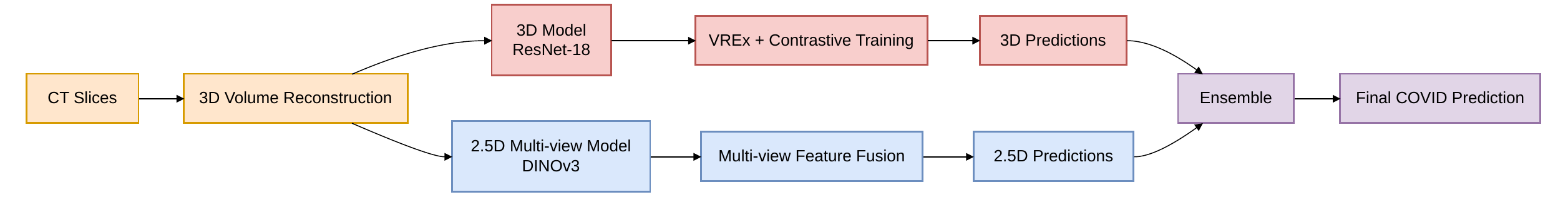}
    \caption{Overview of the proposed method. Axial CT slices are first reconstructed into a normalized $128\times128\times128$ volume through preprocessing. From the reconstructed volume, two complementary representations are learned. The 3D branch processes the full volume using a ResNet-18 architecture trained with Variance Risk Extrapolation (VREx) and supervised contrastive learning to improve cross-domain robustness. In parallel, a 2.5D multi-view branch extracts axial, coronal, and sagittal slices and processes them using a DINOv3 backbone. The predictions from both models are aggregated through an ensemble to obtain the final classification.}
    \label{fig:method}
\end{figure*}

\subsection{Data Preprocessing}

The raw datasets consist of axial CT slices. To obtain consistent volumetric inputs, we reconstruct full 3D CT volumes by stacking slices along the axial axis. Duplicate slices are removed prior to reconstruction.

Each reconstructed scan is resized to a fixed spatial resolution of $128 \times 128 \times 128$. We then apply a 3D Gaussian denoising filter followed by mask sharpening to improve anatomical clarity. Finally, voxel intensities are normalized into grayscale values in the range $[0,255]$.

\begin{figure}[ht]
    \centering
    \includegraphics[width=\linewidth]{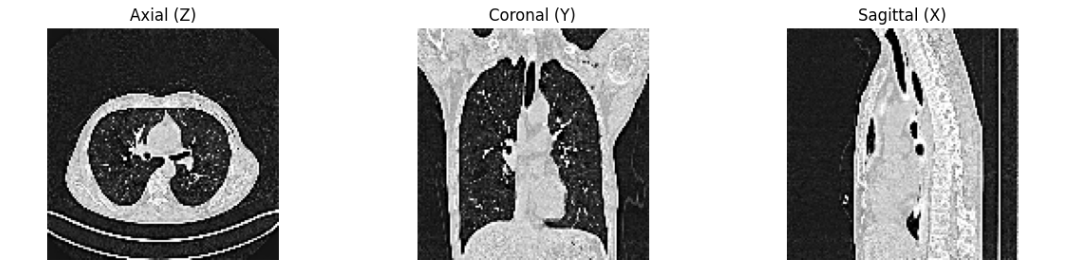}
    \caption{From each reconstructed CT volume, we extract slices from three orthogonal anatomical planes: Axial view (original acquisition plane), coronal view, sagittal view.}
    \label{fig:ct_slices}
\end{figure}

\subsection{3D Representation Learning}

To model volumetric context, we adopt a 3D convolutional architecture based on ResNet-18. The model is adapted for single-channel CT input by modifying the first convolutional layer.

\subsubsection{Architecture}

The 3D branch uses a ResNet-18 architecture designed for video processing~\cite{hara3dresnet}. The network receives $128 \times 128 \times 128$ volumes as input and produces feature embeddings that are passed to a task-specific classification head.

\subsubsection{Training Strategy}

\paragraph{Stage 1: Domain Generalization Pretraining}

The model is first pretrained using Variance Risk Extrapolation (VREx) \cite{pmlr-v139-krueger21a}, a domain generalization technique that encourages consistent performance across different data distributions. The objective combines cross-entropy classification loss with a variance regularization term computed across domain-specific losses. During this stage, auxiliary domain labels (e.g., data source or demographic attributes) are used to define domains. 

\paragraph{Stage 2: Task-Specific Fine-tuning}

After domain pretraining, the network is fine-tuned for the downstream classification task. The training objective combines cross-entropy loss with supervised contrastive learning \cite{khosla2020supervised} to improve feature separability. In addition, MixUp augmentation \cite{zhang2018mixup} is applied to encourage smoother decision boundaries and improve generalization.

\subsection{2.5D Multi-View Representation Learning}

While 3D models capture global structure, they can be computationally expensive and may overlook fine-grained slice-level patterns. To complement the 3D approach, we introduce a 2.5D multi-view framework based on transformer-based visual representations.

\subsubsection{Multi-View Slice Extraction}

From each reconstructed CT volume, we extract slices from three orthogonal anatomical planes: Axial view (original acquisition plane), coronal view, sagittal view. Representative slices are sampled from each plane to capture relevant lung structures. These slices are resized to $224 \times 224$ resolution before being processed by the model.

\subsubsection{DINOv3 Backbone}

For feature extraction, we adopt a Vision Transformer backbone pretrained with the DINOv3 self-supervised learning framework \cite{simeoni2025dinov3}. The pretrained model provides strong visual representations that transfer effectively to medical imaging tasks. The grayscale CT slices are adapted to the input requirements of the pretrained model, and a lightweight classification head is added on top of the extracted embeddings.

\subsubsection{Multi-View Feature Fusion}

Each view is processed independently through the backbone network. The resulting embeddings are combined using a feature-level fusion strategy. This multi-view aggregation allows the model to capture complementary anatomical information across planes.

\subsection{Ensemble Integration}


We integrate the outputs using an ensemble strategy that aggregates the predicted probabilities from both models. This fusion improves robustness and stabilizes predictions across heterogeneous imaging sources.

%% file: sec/4_experiments.tex
\section{Experiments}
\label{sec:experiments}

\subsection{Datasets}

We evaluate our method on the PHAROS-AIF-MIH benchmark, which focuses on robust disease diagnosis from medical imaging under multi-source and fairness-aware evaluation settings. The dataset consists of chest CT scans collected from multiple institutions, introducing variations in imaging protocols, scanners, and patient populations that create realistic domain shifts. The benchmark includes two tasks:

\begin{itemize}
\item \textbf{Task 1:} Binary COVID-19 detection from chest CT scans. The dataset contains COVID-19 and non-COVID-19 cases collected from multiple sources. The distribution of samples is summarized in Table~\ref{tab:task1_data}.
\item \textbf{Task 2:} Multi-class classification involving four categories: \textit{Healthy}, \textit{Adenocarcinoma}, \textit{Squamous Cell Carcinoma}, and \textit{COVID-19}. This task also evaluates fairness by reporting performance across different gender groups. The distribution of samples is summarized in Table~\ref{tab:task2_data}.
\end{itemize}

These two tasks allow evaluation of both diagnostic performance and model robustness under heterogeneous data sources and demographic variations.

\begin{table}[h]
\centering
\caption{Data samples in the Multi-Source COVID-19 Detection Challenge}
\label{tab:task1_data}
\begin{tabular}{lccc}
\hline
Set & COVID-19 & Non-COVID-19 & Total \\
\hline
Train & 564 & 660 & 1,224 \\
Validation & 128 & 180 & 308 \\
\hline
Total & 692 & 840 & 1,532 \\
\hline
\end{tabular}
\end{table}

\begin{table}[h]
\centering
\small
\caption{Data samples in the Fair Disease Diagnosis Challenge. For Task 2, the multi-class classification involves four distinct diagnostic categories: Healthy, Adenocarcinoma, Squamous Cell Carcinoma, and COVID-19. For brevity and visual clarity in our subsequent tables and figures, we map these clinical categories to the following abbreviations: \textbf{`Normal'} (Healthy), \textbf{`A'} (Adenocarcinoma), \textbf{`G'} (Squamous Cell Carcinoma), and \textbf{`COVID'} (COVID-19).}
\label{tab:task2_data}
\setlength{\tabcolsep}{4pt}
\begin{tabular}{lcccc}
\hline
Set & Healthy & \makecell{Adeno-\\carcinoma} & \makecell{Squamous\\Cell} & COVID \\
\hline
Train (F) & 100 & 125 & 5 & 100 \\
Train (M) & 100 & 125 & 79 & 100 \\
Val (F) & 20 & 25 & 13 & 20 \\
Val (M) & 20 & 25 & 12 & 20 \\
\hline
Total (F) & 120 & 150 & 18 & 120 \\
Total (M) & 120 & 150 & 91 & 120 \\
\hline
\end{tabular}
\end{table}

\subsection{Implementation Details}

All models are implemented in PyTorch~\cite{paszke2019pytorch} and trained on NVIDIA GPUs using CUDA with automatic mixed precision (AMP) to improve training efficiency and reduce memory usage.

\paragraph{3D Model.}
Training follows a two-stage procedure. In the first stage, the model is pretrained using Variance Risk Extrapolation (VREx) to encourage domain-invariant representations across sources. The loss combines cross-entropy with a variance penalty across domain losses, with regularization weight $\lambda=1.0$. This stage is trained for 5 epochs using AdamW with a learning rate of $1\times10^{-4}$.

In the second stage, the model is fine-tuned for classification using cross-entropy combined with supervised contrastive learning ($\tau=0.07$). MixUp augmentation with $\alpha=0.4$ is applied to improve robustness. Fine-tuning runs for 20 epochs with a learning rate of $1\times10^{-5}$ and cosine annealing scheduling. Weight decay is set to $1\times10^{-5}$.

\paragraph{2.5D Multi-View Model.}
For each CT volume, slices are extracted from axial, coronal, and sagittal planes, with 8–12 slices uniformly sampled per view. Each slice is processed independently using a DINOv3 backbone, and the resulting embeddings are projected and fused across views.

Training proceeds in three stages: (1) training the classification head with the backbone frozen for 10 epochs ($1\times10^{-3}$ learning rate), (2) progressively unfreezing upper transformer layers for 15 epochs ($1\times10^{-4}$), and (3) full end-to-end fine-tuning for 20 epochs ($5\times10^{-5}$). Optimization uses AdamW with cosine learning rate scheduling and a 5\% warmup.

\paragraph{Data Augmentation.}
Spatial augmentations include random rotations ($\pm15^\circ$), horizontal flips (0.5), and scaling (0.8–1.2). Intensity augmentations include brightness and contrast adjustments, Gaussian noise ($\sigma=0.01$), and random cutout with 10\% masked area. Augmentations are applied consistently across slices within each view.

\subsection{Training Protocol}

The 3D model is trained using the two-stage procedure described earlier: domain generalization pretraining followed by supervised fine-tuning. 

For the 2.5D model, training begins with the backbone frozen while optimizing the classification head. Transformer layers are progressively unfrozen in later stages to adapt pretrained representations to the medical imaging domain. Standard spatial and intensity augmentations are applied to improve generalization.

\subsection{Evaluation Protocol}

Performance is evaluated on the validation sets provided by the challenge organizers. Following the benchmark guidelines, we report overall predictive performance as well as robustness across data sources and demographic groups.

To further improve reliability, predictions from the 2.5D and 3D branches are combined through ensemble inference.

\subsection{Results}

\subsubsection{3D ResNet-18 Results}

\paragraph{Task 1.}
For the binary COVID-19 detection task, the 3D model achieves a validation accuracy of \textbf{87.01\%} and a mean Macro F1-score of \textbf{0.7648} across the 4 sources. Table~\ref{tab:task1_source_results} reports the performance across different data sources. While the model performs strongly on most sources, performance degradation is observed on Source~2, highlighting the impact of domain shift across institutions.

\begin{table}[h]
\centering
\caption{Per-source performance for Task 1 using the 3D ResNet-18 model.}
\label{tab:task1_source_results}
\begin{tabular}{lcccc}
\hline
Source & Source 0 & Source 1 & Source 2 & Source 3 \\
\hline
F1-score & 0.8630 & 0.8408 & 0.4828 & 0.8725 \\
\hline
\end{tabular}
\end{table}

\paragraph{Task 2.}
For the multi-class classification task (A, G, COVID, Normal), the 3D model achieves an overall validation accuracy of \textbf{76.77\%} and a mean Macro F1-score of \textbf{0.6677} between the 2 genders. Table~\ref{tab:task2_class_results} shows the class-wise performance. 

To assess fairness across demographic groups, we analyze performance across gender. The model achieves an F1-score of \textbf{0.7249} for male patients and \textbf{0.6104} for female patients, indicating a moderate performance gap that highlights the importance of fairness-aware evaluation.

\begin{table}[h]
\centering
\caption{Class-wise performance for Task 2 using the 3D ResNet-18 model.}
\label{tab:task2_class_results}
\begin{tabular}{lccc}
\hline
Class & Precision & Recall & F1-score \\
\hline
A & 0.6901 & 0.9800 & 0.8099 \\
G & 0.7500 & 0.1200 & 0.2069 \\
COVID & 0.8462 & 0.8250 & 0.8354 \\
Normal & 0.8293 & 0.8500 & 0.8395 \\
\hline
\end{tabular}
\end{table}

\subsubsection{2.5D DINOv3 Results}

\paragraph{Task 1.}
We evaluate the 2.5D multi-view model based on the DINOv3 backbone on the binary COVID-19 detection task. The model achieves a validation accuracy of \textbf{93.51\%} and a mean Macro F1-score of \textbf{0.8221}, demonstrating strong performance using multi-view slice representations. Table~\ref{tab:task1_dino_results} shows the per-source F1-score across the four data sources.

\begin{table}[h]
\centering
\caption{Per-source performance for Task 1 using the 2.5D DINOv3 model.}
\label{tab:task1_dino_results}
\begin{tabular}{lcccc}
\hline
Source & Source 0 & Source 1 & Source 2 & Source3\\
\hline
F1-score & 0.9430 & 0.9431 & 0.4828 & 0.9194 \\
\hline
\end{tabular}
\end{table}

\paragraph{Task 2.}

For the multi-class classification task (A, G, COVID, Normal), the 2.5D DINOv3-based model achieves an overall validation accuracy of \textbf{76.77\%} and a mean Macro F1-score of \textbf{0.7229}. Table~\ref{tab:task2_dino_results} presents the class-wise performance of the model.

To assess fairness across demographic groups, we analyze performance across gender. The model achieves an F1-score of \textbf{0.7848} for male and \textbf{0.6611} for female patients. The performance difference suggests the presence of domain shifts between institutions, which remains a challenge for multi-source medical imaging models.

\begin{table}[h]
\centering
\caption{Class-wise performance for Task 2 using the 2.5D DINOv3 model.}
\label{tab:task2_dino_results}
\begin{tabular}{lccc}
\hline
Class & Precision & Recall & F1-score \\
\hline
A & 0.7241 & 0.8400 & 0.7778 \\
G & 0.5625 & 0.3600 & 0.4390 \\
COVID & 0.8333 & 0.8750 & 0.8537 \\
Normal & 0.8462 & 0.8250 & 0.8354 \\
\hline
\end{tabular}
\end{table}

\subsubsection{Ensemble Results}

Finally, we evaluate an ensemble that combines the predictions of the 3D ResNet-18 and the 2.5D DINOv3 models. The ensemble is implemented by averaging the logits from both models and selecting the final prediction based on the combined scores. This strategy allows the framework to leverage complementary information captured by volumetric representations and multi-view slice features.

\paragraph{Task 1.}

The ensemble model achieves the best overall performance, reaching a validation accuracy of \textbf{94.48\%} and a mean Macro F1-score of \textbf{0.9426}. Table~\ref{tab:task1_model_comparison} summarizes the comparison between the models and the ensemble.

The ensemble improves performance across most sources, particularly on Source~0 and Source~1, demonstrating the benefit of combining volumetric and slice-based representations.

\begin{table}[h]
\centering
\caption{Performance comparison for Task 1 (COVID-19 detection). Best results are shown in bold.}
\label{tab:task1_model_comparison}
\begin{tabular}{lcc}
\hline
Model & Accuracy & Macro F1 \\
\hline
3D ResNet-18 & 0.8701 & 0.7648 \\
2.5D DINOv3 & 0.9351 & 0.8221 \\
Ensemble & \textbf{0.9513} & \textbf{0.8321} \\
\hline
\end{tabular}
\end{table}

\begin{table}[h]
\centering
\caption{Per-source F1 comparison for Task 1. Best values are shown in bold.}
\label{tab:task1_source_comparison}
\begin{tabular}{lcccc}
\hline
Model & Source 0 & Source 1 & Source 2 & Source 3 \\
\hline
3D ResNet-18 & 0.8630 & 0.8408 & 0.4828 & 0.8725 \\
2.5D DINOv3 & 0.9430 & \textbf{0.9431} & 0.4828 & \textbf{0.9194} \\
Ensemble & \textbf{0.9659} & \textbf{0.9431} & \textbf{0.5000} & \textbf{0.9194} \\
\hline
\end{tabular}
\end{table}

\paragraph{Task 2.}

For the multi-class disease classification task, the ensemble achieves an overall validation accuracy of \textbf{76.77\%} and a mean Macro F1-score of \textbf{0.7229}. While the ensemble improves stability across sources, the best overall performance for this task is still achieved by the 2.5D DINOv3 model.

Table~\ref{tab:task2_model_comparison} compares the performance of the models and the ensemble.

\begin{table}[h]
\centering
\caption{Performance comparison for Task 2 (multi-class classification). Best results are shown in bold.}
\label{tab:task2_model_comparison}
\begin{tabular}{lcc}
\hline
Model & Accuracy & Macro F1 \\
\hline
3D ResNet-18 & \textbf{0.7677} & 0.6677 \\
2.5D DINOv3 & \textbf{0.7677} & \textbf{0.7230} \\
Ensemble & \textbf{0.7677} & {0.7229} \\
\hline
\end{tabular}
\end{table}

\begin{table}[h]
\centering
\caption{Class-wise F1 comparison for Task 2. Best values are shown in bold.}
\label{tab:task2_class_comparison}
\begin{tabular}{lccc}
\hline
Class & 3D ResNet-18 & 2.5D DINOv3 & Ensemble \\
\hline
A & 0.8099 & \textbf{0.7778} & \textbf{0.7778} \\
G & 0.2069 & \textbf{0.4390} & \textbf{0.4390} \\
COVID & 0.8354 & \textbf{0.8537} & \textbf{0.8537} \\
Normal & 0.8395 & \textbf{0.8537} & 0.8354 \\
\hline
\end{tabular}
\end{table}

\subsection{Test Submission Results}

Our methods were evaluated on the official test server of the PHAROS-AIF-MIH benchmark. The proposed approach achieved strong rankings across both tasks, demonstrating robustness to multi-source distribution shifts and fairness considerations.

\paragraph{Multi-Source COVID-19 Detection (Task 1).}
Our team achieved \textbf{runner-up} position in the challenge. Table~\ref{tab:task1_test} summarizes the performance across different hospitals. The 2.5D model outperforms the standalone 3D model, while ensembling further improves robustness. The best performance is obtained with the 0.5-weighted ensemble, achieving a Macro F1-score of \textbf{0.751}.

\begin{table}[h]
\centering
\caption{Test results for Task 1 across hospitals.}
\label{tab:task1_test}
\begin{tabular}{lccccc}
\hline
Method & Avg & H$_1$ & H$_2$ & H$_3$ & H$_4$ \\
\hline
2.5D & 0.741 & 0.910 & 0.691 & 0.493 & 0.869 \\
3D & 0.699 & 0.825 & 0.608 & 0.495 & 0.868 \\
Ens (0.3) & 0.739 & 0.895 & 0.673 & 0.496 & 0.892 \\
Ens (0.5) & \textbf{0.751} & \textbf{0.914} & \textbf{0.712} & 0.495 & 0.883 \\
Ens (0.7) & 0.744 & 0.912 & 0.696 & 0.495 & 0.873 \\
\hline
\end{tabular}
\end{table}

The results show that ensembling improves generalization across most hospitals, particularly for H$_2$, while performance on H$_3$ remains challenging for all methods.

\paragraph{Fair Disease Diagnosis (Task 2).}
Our team ranked \textbf{third place} in this task. Table~\ref{tab:task2_test} reports the Macro F1-scores along with gender-specific performance. The best result is achieved by the 0.7-weighted ensemble, with a Macro F1-score of \textbf{0.633}.

\begin{table}[h]
\centering
\caption{Test results for Task 2 with gender-based evaluation.}
\label{tab:task2_test}
\begin{tabular}{lccc}
\hline
Method & Avg & Female & Male \\
\hline
2.5D & 0.555 & 0.709 & 0.402 \\
3D & 0.572 & \textbf{0.756} & 0.388 \\
Ens (0.3) & 0.568 & 0.732 & 0.403 \\
Ens (0.5) & 0.561 & 0.718 & 0.403 \\
Ens (0.7) & \textbf{0.633} & 0.709 & \textbf{0.557} \\
\hline
\end{tabular}
\end{table}

The results highlight a noticeable performance gap between genders, with higher performance on female samples for most methods. However, the 0.7-weighted ensemble significantly improves male performance, leading to the best overall Macro F1-score. This suggests that adjusting ensemble weights can help balance fairness and overall accuracy.

\subsection{Discussion}

The experimental results highlight the complementary strengths of volumetric and slice-based representations for CT analysis. The 2.5D DINOv3 model consistently outperforms the 3D ResNet-18 model across both tasks, suggesting that large-scale pretrained visual representations provide strong feature extraction capabilities even for medical imaging data. In particular, the transformer-based architecture benefits from pretraining on diverse visual datasets, enabling improved discrimination of subtle patterns in CT slices.

The ensemble further improves performance for the binary COVID-19 detection task, achieving the highest overall accuracy and Macro F1-score. This improvement indicates that the 3D and 2.5D branches capture complementary information: the 3D model captures global volumetric context, while the 2.5D model focuses on high-resolution slice-level features. Combining these representations allows the ensemble to leverage both global and local anatomical cues.

However, the ensemble does not improve performance for the multi-class disease classification task, where the 2.5D model already achieves the best results. This suggests that slice-based features may be sufficient for distinguishing disease categories, while volumetric context provides limited additional benefit for this task.

While the ensemble demonstrates robust generalization across most data sources, we observe a severe performance degradation in distinguishing Squamous Cell Carcinoma (Class G) from other pathologies severely exacerbates these errors. This suggests that while global 3D volumetric features capture excellent broader context, they are significantly less resilient to severe spatial distribution shifts than 2.5D slice-level representations.

%% file: sec/5_conclusion.tex
\section{Conclusion}
\label{sec:conclusion}

In this paper, we proposed a hybrid deep learning framework that integrates 2.5D slice-level features via a DINOv3 vision transformer and 3D volumetric context via a ResNet-18 architecture for robust COVID-19 detection and multi-class severity classification. Rather than relying on a single dimensional paradigm, our approach successfully leverages the complementary strengths of both. 

Evaluations on the multi-source PHAROS-AIF-MIH benchmark reveal a clear functional dichotomy: volumetric 3D models and their ensembles excel at extracting global context for binary classification tasks (effectively distinguishing COVID-19 from non-COVID cases), whereas slice-based 2.5D features demonstrate superior standalone efficacy for complex, multi-class pathology grading. By combining these representations through logit-level ensembling and stabilizing the feature space with VREx and contrastive learning, our framework yields a highly resilient diagnostic tool. Ultimately, this work offers a practical pathway toward clinically viable, cross-source medical image analysis that can withstand the domain shifts inherently found in real-world hospital environments.